\documentclass[journal]{IEEEtran}
\input{decl}

\begin{document}

\title{POCS-based framework of signal reconstruction from generalized non-uniform samples}

\author{Nguyen T. Thao,~\IEEEmembership{Member,~IEEE,}
Dominik Rzepka,~\IEEEmembership{Member,~IEEE,} and Marek Mi\'skowicz,~\IEEEmembership{Senior Member,~IEEE} % <-this % stops a space
\thanks{N. T. Thao is with the Department of Electrical Engineering, The City College of New York, CUNY, New York, USA, email: tnguyen@ccny.cuny.edu.}% <-this % stops a space
\thanks{D. Rzepka and M. Mi\'skowicz are with the Department of Measurement and Electronics, AGH University of Science and Technology, Krak\'ow, Poland, emails: drzepka@agh.edu.pl, miskow@agh.edu.pl}% <-this % stops a space
\thanks{D. Rzepka and M. Mi\'skowicz were supported by the Polish National Center of Science under grants DEC-2017/27/B/ST7/03082 and DEC-2018/31/B/ST7/03874, respectively.}
}
\maketitle

\begin{abstract}
We formalize the use of projections onto convex sets (POCS) for the reconstruction of signals from non-uniform samples in their highest generality. This covers signals in any Hilbert space $\scH$, including multi-dimensional and multi-channel signals, and samples that are most generally inner products of the signals with given kernel functions in $\scH$.  An attractive feature of the POCS method is the {\em unconditional} convergence of its iterates to an estimate that is consistent with the samples of the input, even when these samples are of very heterogeneous nature on top of their non-uniformity, and/or under insufficient sampling. Moreover, the error of the iterates is systematically monotonically decreasing, and their limit retrieves the input signal whenever the samples are uniquely characteristic of this signal. In the second part of the paper, we focus on the case where the sampling kernel functions are orthogonal in $\scH$, while the input may be confined in a smaller closed space $\scA$ (of bandlimitation for example). This covers the increasingly popular application of time encoding by integration, including multi-channel encoding. We push the analysis of the POCS method in this case by giving a special parallelized version of it, showing its connection with the pseudo-inversion of the linear operator defined by the samples, and giving a multiplierless discrete-time implementation of it that paradoxically accelerates the convergence of the iteration.

\end{abstract}

\section{Introduction}

The reconstruction of bandlimited signals from non-uniform samples is a difficult topic that has been studied since the 50's \cite{Duffin52,Yen56,Benedetto92,Feichtinger94,Marvasti01}, although its practical development has remained somewhat limited. But this subject is currently attracting new attention with the increasing trend of
event-based sampling in data acquisition \cite{Miskowicz2018,Rzepka18,Alexandru20,Adam21,Tarnopolsky22}. This approach to sampling has grown in an effort to simplify the complexity of the analog sampling circuits, lower their power consumption and simultaneously increase their precision. This is made possible in particular by the replacement of amplitude encoding by time encoding, which takes advantage of the higher precision of solid-state circuits in time measurement. Time encoding has however induced the use of non-uniform samples that were not commonly studied in the past literature. A well-known example is the time encoder of \cite{Lazar04} which acquires the integrals of an input signal over successive non-uniform intervals and makes use of a special algorithm introduced in \cite{Feichtinger94} for signal reconstruction. While a breakthrough in time encoding, this invention has however the shortcomings of strict conditions for the convergence of the algorithm and a limited potential for generalizations to other types of non-uniform sampling such as, for example, leaky integrate-and-fire encoding \cite{LAZAR2005401} (as pointed in \cite{Thao22a}).

\begin{figure}
\centerline{\hbox{\scalebox{0.9}{\includegraphics{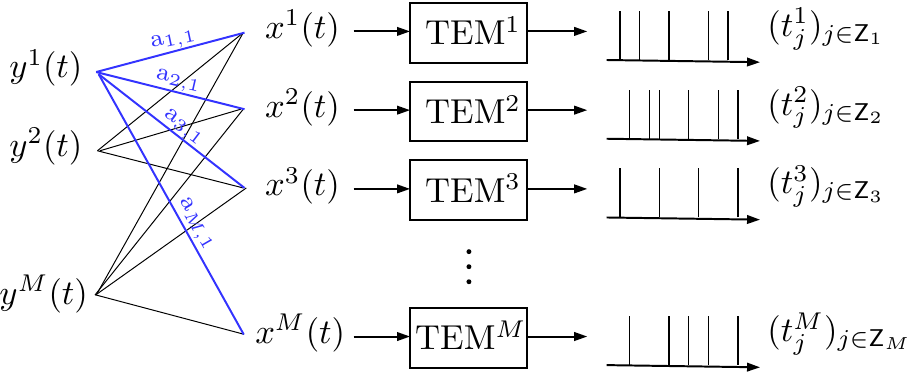}}}}
\caption{Multi-channel system of  time-encoding machines (TEM) from \cite{Adam20,Adam21}.}\label{fig:Karen}
\end{figure}

The method of projection onto convex sets (POCS) was initially introduced in \cite{Yeh90} for signal reconstruction from non-uniform point samples.  But it was later deemed a slow method in \cite{Feichtinger94} and since then has not retained much attention in sampling. Interest in this method however got more recently revived with the following features and events: 1) the time-encoding reconstruction algorithm of \cite{Lazar04} was improved in \cite{Thao21a} by an adaptation of the POCS algorithm with a similar rate of convergence but {\em unconditional} convergence and a lower computation complexity; 2) the inherent versatility of the POCS method makes it an attractive candidate in the current trend of event-based sampling, which is pushing the use of non-standard sampling schemes of various nature and complexity; 3)  there exists a pool of POCS techniques that has been developed over decades \cite{Combettes93,Bauschke96,combettes1997hilbertian} but has not been extensively exploited in non-uniform sampling, while the Kaczmarz method \cite{Kaczmarz37} as a particular case of the POCS algorithm is independently attracting attention in big data  \cite{strohmer2009randomized,Needell14,Ma15}.  An example of versatility of the POCS method is its application in \cite{Adam20,Adam21} to the multi-channel time encoding system shown in Fig. \ref{fig:Karen}.

With this recent comeback of the POCS method in nonuniform sampling, the purpose of this article is to take some distance and reflect on what this method can fundamentally achieve in this problem at the highest possible level of generality.  This method is applicable in any abstract Hilbert space $\scH$, separable or not. Whenever an input signal $x\in\scH$ is known to be in the intersection of closed convex subsets of $\scH$, the POCS method can systematically find an element in this intersection as an estimate of $x$.  This is the case when an encoder extracts from $x$ a sequence of samples $(\s_k)_{k\in\Z}$ of the form
\begin{equation}\label{sample}
\s_k:=\langle x,h_k\rangle,\qquad k\in\Z
\end{equation}
where  $\langle\cdot,\cdot\rangle$ is the inner product of $\scH$ and $(h_k)_{k\in\Z}$ is a known collection of functions in $\scH$. Indeed, this tells us that $x$ belongs to the intersection of the affine hyperplanes $\scC_k:=\{v\in\scH:\langle v,h_k\rangle=\s_k\}$, which are a particular case of closed convex sets. When $x$ is  additionally known to be in a closed subspace $\scA$ of $\scH$ of ``limited freedom'', such as bandlimitation, this subspace is naturally to be incorporated by the POCS method among the convex sets. In the basic sampling situation where $x$ is a bandlimited function $x(t)$ and $\s_k=x(t_k)$ at some instant $t_k$, then $\s_k$ does yield the form of \eqref{sample} where $h_k(t)$ is a sinc function shifted by $t_k$. But as \eqref{sample} refers to a more general situation, we call $\langle x,h_k\rangle$ a {\em generalized sample} of $x$ \cite{eldar2005general} and $h_k$ the corresponding {\em sampling kernel function}.
We give in Section \ref{sec:gen-samp} an overview of concrete non-uniform sampling examples that are covered by the formalism of \eqref{sample}. This includes point sampling, derivative sampling, integrate-and-fire encoding, asynchronous Sigma-Delta modulation (ASDM) with bandlimited signals in $L^2(\RR)$ or more generally $L^2(\RR^N)$ or $(L^2(\RR))^N$. In the remainder of the paper, we then elaborate on the specific contributions of the POCS method in the general setting of \eqref{sample}.

Section \ref{sec:gen-POCS} is an overview of the basic principles of the POCS method and its various versions when applied to \eqref{sample} with $x$ in a closed subspace $\scA$ as only extra assumption. When $\Z$ is finite, which is always the case in practice, an outstanding property of the POCS method is the {\em systematic} convergence of its iteration, contrary to the algorithm of \cite{Lazar04}, whether perfect reconstruction is theoretically possible or not. More specifically, it tends to the estimate that is closest to the initial iterate while being consistent with the samples of $x$. This automatically leads to perfect reconstruction whenever the samples are uniquely characteristic of $x$, whether one is able to prove this situation or not. Meanwhile, there is no restriction on the type of samples that can be used. One can for example mix samples from derivatives of the input of various orders together with integral values. For illustration, we show an experiment of perfect reconstruction from input extrema whose density is below the Nyquist rate. This is possible as the 0-derivative property of extrema has the effect to double the number of samples in the generalized sense.

In the second part of the paper, we focus on the special case where the sampling kernel functions $(h_k)_{k\in\Z}$ are orthogonal in $\scH$ while $x$ remains in the smaller space $\scA$. This is a case of importance as this includes the time-encoding machine of \cite{Lazar04}, leaky integrate-and-fire encoding \cite{LAZAR2005401}, as well as multi-channel time encoders such as those introduced in \cite{Adam20b,Adam20,Adam21}. This special case turns out to exhibit a number of outstanding properties, that were initially noted in \cite{Thao21a,Thao22a} for the time-encoding scheme of \cite{Lazar04}. These properties include a special parallelized version of the POCS method, its unconditional convergence even when $\Z$ is infinite, its fundamental connection to the Moore-Penrose pseudo-inversion of the linear operator $u\in\scA\mapsto(\langle u,h_k\rangle)\inZ$  (after some normalization), a rigorous and simple DSP discretization of the iterative part of the method that is applicable even when the subspace $\scA$ is of uncountable infinite dimension (case of non-separable Hilbert space), and a multiplierless variant of it that paradoxically accelerates the convergence. This is covered in Sections \ref{sec:ortho-POCS}, \ref{sec:multi-TEM} and \ref{sec:bin}. In Section \ref{sec:multi-TEM} in particular, we show how the multi-channel time-encoder of  \cite{Adam20,Adam21} concisely fits in this framework, thus inheriting all of the above properties and features.

\section{Generalized sampling examples}\label{sec:gen-samp}

\subsection{Basic point sampling in $L^2(\RR)$}

In the basic framework of sampling, $\scH$ is the space $L^2(\RR)$ of square-integrable functions equipped with the inner product
$$\langle u,v\rangle:=\int_\RR u(t)v(t)\,\dif t,\qquad u(t),v(t)\in L^2(\RR),$$
and $\scA$ is a space of bandlimited signals. Up to a change of time unit, one can always assume that $\scA$ is equal to the space $\scB$ of bandlimited signals of Nyquist period 1. Then the most basic example of sampling consists in acquiring from an input $x(t)\in\scB$ the scalar values
\begin{equation}\label{point-samples}
\s_k:=x(t_k),\qquad k\in\Z
\end{equation}
where $\Z$ is some set of consecutive integers, and $(t_k)\inZ$ is some increasing sequence of instants. In the present context of non-uniform sampling, these instants are not assumed to be equidistant. Since $x(t)=\sinc(t)*x(t)$ where $*$ is the convolution product and
$$\sinc(t):=\sin(\pi t)/(\pi t),$$
we also have $\s_k=(\sinc*x)(t_k)=\int_\RR\sinc(t_k-t)x(t)\,\dif t$. Then $\s_k$ can be put in the form of \eqref{sample} with
$$h_k(t):=\sinc(t_k\!-t)=\sinc(t-t_k),\qquad k\in\Z.$$

\subsection{Examples of generalized sampling in $L^2(\RR)$}

One can consider more generally samples of the type
\begin{equation}\label{point-samples-gen}
\s_k:=(f_k*x)(t_k),\qquad k\in\Z
\end{equation}
where $(f_k(t))\inZ$ is some family of functions in $L^2(\RR)$. With some similar derivation, $\s_k$ yields the form of \eqref{sample} with
\begin{equation}\label{hf}
h_k(t):=f_k(t_k\!-t),\qquad k\in\Z.
\end{equation}
In this generalization, we see two generic examples.

\subsubsection{Derivative sampling}

This is the case where derivatives of $x(t)$ of arbitrary orders are sampled, giving a sequence of the type
\begin{equation}\label{deriv-samp}
\s_k:=\frac{\dif^{n_k}x}{\dif t^{n_k}}(t_k),\qquad k\in\Z
\end{equation}
where $(n_k)\inZ$ is some sequence of non-negative integers and $(t_k)\inZ$ may be more generally monotonically increasing (allowing for example $t_{k+1}=t_k$ with $n_{k+1}\neq n_k$). This is the particular case of \eqref{point-samples-gen} where
$$f_k(t):=\frac{\dif^{n_k}\sinc}{\dif t^{n_k}}(t),\qquad k\in\Z.$$
Given the even symmetry of $\sinc(t)$, the resulting function $h_k(t)$ of \eqref{hf} is
\begin{equation}\label{hk-deriv}
h_k(t)=(-1)^{n_k}\,\frac{\dif^{n_k}\sinc}{\dif t^{n_k}}(t-t_k),\qquad k\in\Z.
\end{equation}
The samples of \eqref{point-samples} are themselves the particular case of \eqref{deriv-samp} where $n_k=0$ for all $k\in\Z$.

\subsubsection{Integrate-and-fire sampling}\label{subsub:IFS}

In this case, the samples are of the type
\begin{equation}\label{gen-IFS}
\s_k:=\int_{t_{k-1}}^{t_k}f(t_k\!-t)\,x(t)\,\dif t,\qquad k\in\Z
\end{equation}
where $f(t)$ is some function in $L^2(\RR)$. This is the particular case of \eqref{point-samples-gen} where
$$f_k(t):=\left\{\begin{array}{ccl}
f(t)&,&0\leq t<t_k\!-t_{k-1}\\
0&,&\mbox{otherwise}\end{array}\right..$$
Meanwhile, one can see directly from \eqref{gen-IFS} that $\s_k$ yields the form of \eqref{sample} with
\begin{equation}\label{hk-int}
h_k(t):=\left\{\begin{array}{ccl}
f(t_k\!-t)&,&t\in[t_{k-1},t_k)\\
0&,&\mbox{otherwise}\end{array}\right..
\end{equation}
A well-known case is leaky integrate-and-fire encoding (LIF) where
$$f(t):=e^{-\alpha t}$$
for some constant $\alpha\geq0$. In the case $\alpha=0$, the samples are of the simple form
\begin{equation}\label{ASDM}
\s_k:=\int_{t_{k-1}}^{t_k}x(t)\,\dif t,\qquad k\in\Z
\end{equation}
which is also the type of samples that one can extract from an asynchronous Sigma-Delta modulator (ASDM) \cite{Lazar04,Thao21a}.

\subsection{Examples with more sophisticated signal spaces}

\subsubsection{Multi-dimensional signals}

This is the case where $\scH=L^2(\RR^N)$. Each element $x$ in this space is a scalar function $x(\bt)$ where $\bt\in\RR^N$. The space $\scA$ is then a subspace of bandlimited functions in the sense of the $N$-dimensional Fourier transform. The case $N=2$ has been of particular interest for image processing \cite{chen1987analysis,zakhor1990reconstruction} but has been limited to point samples $\s_k=x(\bt_k)$ where $(\bt_k)\inZ$ is a sequence of $\RR^N$.

\subsubsection{Multi-channel signals}

This applies to the system shown in Fig. \ref{fig:Karen} and introduced in \cite{Adam20,Adam21}. Formally in this case, $\scH=(L^2(\RR))^M$ and $\scA$ is a subspace of $\scB^M$. We will study this example in depth in Section \ref{sec:multi-TEM}.

\section{POCS method}\label{sec:gen-POCS}

We now return to the general sampling setting of the introduction and give an overview of how the POCS method can be used to reconstruct or estimate an input signal $x$ in a closed subspace $\scA$ of $\scH$ from samples of the type \eqref{sample}.

\subsection{Consistent estimates}

\begin{figure}
\vspace{-5mm}
\centerline{\hbox{\scalebox{1.2}{\includegraphics{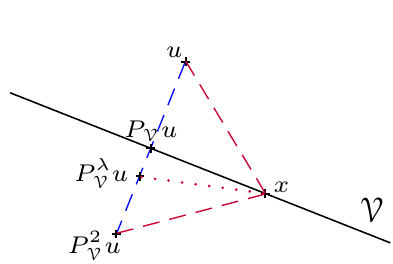}}}}
\caption{Illustration of $P_\scV u$ and $P_\scV^\lambda u$ for $\lambda\in[1,2]$.}\label{fig:proj}
\end{figure}

Looking for an estimate $u$ of $x$, one wishes to make sure that $u$ is at the least consistent with the samples of $x$, meaning that $\langle u,h_k\rangle=\s_k$ for all $k\in\Z$. This is a natural requirement when the samples $(\s_k)\inZ$ are uniquely characteristic of $x$. But there is also a rational reason for doing so when the solution to \eqref{sample} is not unique. From a set theoretic viewpoint, $u$ is consistent if and only if it is in the set
\begin{equation}\label{int-Sk}
\scS:=\textstyle\bigcap\limits\inZ\scS_k\quad\mbox{where}\quad
\scS_k:=\big\{v\in\scA:\langle v,h_k\rangle=\s_k\big\},~~k\in\Z.
\end{equation}
Now, because $\scS$ is the intersection of affine hyperplanes of $\scA$, it is a closed affine subspace of $\scA$ (meaning the translated version of a closed linear subspace of $\scA$). As $\scS$ contains $x$, it then follows from the Pythagorean theorem that the distance of $u$ to $x$ is automatically reduced by projecting $u$ orthogonally to $\scS$ (see Fig. \ref{fig:proj} with $\scV=\scS$). Denoting the norm of $\scH$ by $\|\cdot\|$ and calling $P_\scV$ the orthogonal projection onto any given closed affine subspace $\scV$, we have formally
\begin{equation}\label{error-reduc}
x\in\scV~~\mbox{and}~~ u\notin\scV\quad\Rightarrow\quad\|P_\scV u-x\|<\|u-x\|.
\end{equation}
So if $u\notin\scS$, there is at least theoretical knowledge to improve it as an estimate of $x$. Meanwhile, if $u\in\scS$, there is no more deterministic knowledge to discriminate it from $x$.

In practice, there is unfortunately no closed-form expression for $P_\scS$. However, if $u\notin\scS$, there exists at least some $k\in\Z$ for which $u\notin\scS_k$. Then \eqref{error-reduc} is valid for $\scV:=\scS_k$. Thus, $P_{\scS_k}$ automatically reduces the estimation error of $u$. This time, $P_{\scS_k}$ yields the following simple expression.
\line
\begin{proposition}\label{prop:Pk}
For all $u\in\scH$,
\begin{equation}\label{PSk}
P_{\scS_k}u=P_k u:=\widetilde u+\frac{\s_k-\langle\widetilde u,h_k\rangle}{\|\widetilde h_k\|^2}\,\widetilde h_k
\end{equation}
using the general notation
\begin{equation}\label{tilde-notation}
\widetilde u:=P_\scA u,\qquad u\in\scH.
\end{equation}
\end{proposition}
\line
\begin{IEEEproof}
The following property will be needed,
\begin{equation}\label{basic-eq}
\forall u\in\scH,v\in\scA,\qquad\langle u,v\rangle=\langle\widetilde u,v\rangle
\end{equation}
which is true since $u-\widetilde u$ is by construction orthogonal to $\scA$ and hence to $v$. Let $w$ be the right hand side of \eqref{PSk}. We just need to verify that $w\in\scS_k$ and $u{-}w\perp w{-}v$ for all $v\in\scS_k$. It is first clear that $w\in\scA$. Then $\langle w,h_k\rangle=\langle w,\widetilde h_k\rangle=
\langle\widetilde u,\widetilde h_k\rangle+(\s_k-\langle\widetilde u,h_k\rangle)=\s_k$ as a result of \eqref{basic-eq} again. So $w\in\scS_k$. Next, for any $v\in\scS_k$, $\langle u{-}w,w{-}v\rangle=\langle\widetilde u{-}w,w{-}v\rangle=-\alpha_k\langle\widetilde h_k,w{-}v\rangle$ where $\alpha_k$ is the coefficient of $\widetilde h_k$ in \eqref{PSk}. But $\langle w{-}v,\widetilde h_k\rangle=\langle w{-}v,h_k\rangle=\s_k-\s_k=0$.
\end{IEEEproof}
\ppnoi
Note that in certain cases such as in \eqref{hk-deriv}, $h_k$ is by construction in $\scA$, so that $\widetilde h_k=h_k$. This is however not the case of \eqref{hk-int}.

\subsection{Alternating projections}\label{subsec:alt-proj}

A single projection $P_k u$ is only able to fix the $k$th sample of $u$. To obtain an estimate in $\scS=\cap\inZ\scS_k$, a natural attempt is to perform an iteration of the type
\begin{equation}\label{POCS}
u\up{n+1}=P_{k\up{n}}u\up{n}
\end{equation}
where $(k\up{n})_{n\geq0}$ is some sequence of indices in $\Z$. As $P_\scV u=u$ when $u\in\scV$, what can be claimed from \eqref{error-reduc} is that the estimation error $\|u\up{n}\!-x\|$ will at least monotonically decrease with $n$. If one makes sure that the sets
$$I_k:=\{n\geq0:k\up{n}=k\}$$
are infinite for every $k\in\Z$, then one can further guarantee that $\|u\up{n}\!-x\|$ will stop decreasing only when $u\up{n}$ is effectively in $\scS$.

When $\Z$ is finite, it is known that $u\up{n}$ does eventually converge to an element of $\scS$
\begin{equation}\label{POCS-lim}
\lim_{n\rightarrow\infty}u\up{n}=u\up{\infty}\in\scS
\end{equation}
under the stronger condition that $(k\up{n})_{n\geq0}$ is ``almost cyclic'', which means that the distance between every consecutive elements of $I_k$ is bounded for each $k\in\Z$. This results from the general theory of projections onto convex sets (POCS) \cite{Combettes93,Bauschke96}, affine sets being a particular case of convex sets.
Of course, $u\up{\infty}=x$ when $x$ is the only element of $\scS$. In other words, when the samples $(\s_k)\inZ$ uniquely characterize the input $x$, the POCS iteration of \eqref{POCS} leads to perfect reconstruction. When $\scS$ is not a singleton, it is actually known that
\begin{equation}\label{x-inf}
u\up{\infty}=P_\scS u\up{0}
\end{equation}
because the subspaces $\scS_k$ are affine.
Qualitatively, $u\up{\infty}$ is the element of $\scS$ that is closest to the initial estimate $u\up{0}$ with respect to the norm of $\scH$. If one chooses $u\up{0}=0$, then $u\up{\infty}$ is the minimal norm element of $\scS$.

Because $\scS_k$ is more specifically a hyperplane of $\scA$, the iteration of \eqref{POCS} with a finite set $\Z$ also falls in the description of the Kaczmarz method within the space $\scA$ (note that $u\up{n}\in\scA$ for all $n\geq1$ regardless of $u\up{0}$). The original version of this method is the case where $(k\up{n})_{n\geq0}$ is ``cyclic'' \cite{Kaczmarz37}, which amounts to saying that the elements of $I_k$ are  exactly equidistant of period $\mathrm{card}\,\Z$ for each $k\in\Z$. This was first applied in the basic non-uniform sampling case of \eqref{point-samples} in \cite{Yeh90}. Because of the slow convergence, some other sequences $(k\up{n})_{n\geq0}$ were discussed in \cite{Feichtinger94} in this specific application. Randomly generated sequences $(k\up{n})_{n\geq0}$ were also proposed in \cite{strohmer2009randomized} to accelerate the convergence. But \eqref{POCS-lim} was shown only in a probabilistic sense and under uniqueness of reconstruction.

When $\Z$ is infinite, there is no general result of convergence in the sense of \eqref{POCS-lim}. However, the following weaker convergence
$$\forall k\in\Z,\qquad\lim_{n\rightarrow\infty}\langle u,h_k\up{n}\rangle=\s_k$$
is obtained under an extended version of the ``almost cyclic'' condition (where the bound on the distance between every consecutive elements of $I_k$ may depend on $k\in\Z$) \cite{combettes1997hilbertian}.

\subsection{Relaxed projections}\label{subsec:relax}

The estimation error reduction of \eqref{error-reduc} is in fact more generally realized by a relaxed version of $P_\scV$. For any transformation $P$ and $\lambda\in\RR$, let
\begin{align}
P^\lambda u&:=\lambda Pu+(1-\lambda)u\nonumber\\
&=u+\lambda(Pu-u).\label{rel-P}
\end{align}
Then, a more general version of \eqref{error-reduc} is
\begin{equation}\label{error-reduc-rel}
x\in\scV~\mbox{and}~u\notin\scV~~\Rightarrow~~\forall\lambda\in(0,2),~\|P^\lambda_\scV u-x\|<\|u-x\|.
\end{equation}
This can be seen in Fig. \ref{fig:proj} and is easy to prove from the Pythagorean theorem. Meanwhile, $P^\lambda u$ is still equal to $u$ when $u\in\scV$ for any $\lambda$. The iteration of \eqref{POCS} is then naturally generalized to
\begin{equation}\label{POCS-rel}
u\up{n+1}=P^{\lambda\up{n}}_{k\up{n}}u\up{n}
\end{equation}
where $(\lambda\up{n})_{n\geq}$ is a sequence of relaxation coefficients in $(0,2)$. When these coefficients are more strictly in an interval of the type $[\eps,2{-}\eps]$ for some constant $\eps\in(0,1]$, then all the convergence results of Section \ref{subsec:alt-proj} are known to remain valid. This additional relaxation freedom permits in practice an acceleration of the convergence. There is however little analytical guideline on how to optimally adjust $(\lambda\up{n})_{n\geq0}$.

\subsection{More sophisticated POCS methods}

There exists a large number of more sophisticated POCS methods aiming at accelerating the convergence. A generic technique of interest is the parallel use of multiple projections at each iteration. This involves transformations of the type
\begin{equation}\label{PKmu}
P_K^\bmu u=u+\smallsum{k\in K}\mu_k(P_k u-u)
\end{equation}
where $K$ is some subset of $\Z$ and $\bmu=(\mu_k)_{k\in K}$ is some sequence of coefficients. A basic version is to have positive coefficients $\mu_k$ such that $\sum\inZ\mu_k\in[\eps,2{-}\eps]$. This amounts to having $P_K^\bmu$ equal to a convex combination of individual relaxed projections $P_k^\lambda$ \cite{Bauschke96}.

Whenever possible, a technique to further accelerate the convergence is to decide at each iteration what transformation to apply on $u\up{n}$ depending on its position with respect to the sets $(\scS_k)\inZ$. In the serial case of \eqref{POCS}, a greedy approach is to choose for $k\up{n}$ at each iteration $n$ the index $k$ that minimizes $\|P_k u\up{n}\!-x\|$. By the Pythagorean theorem, this is equivalent to maximizing $\|P_k u\up{n}\!-u\up{n}\|$, which amounts to choosing the set $\scS_k$ that is ``most remote'' from  $u\up{n}$. With the parallel scheme of \eqref{PKmu}, an adaptive approach allows to choose coefficients $\mu_k$ of substantially larger magnitudes and leading to dramatic convergence accelerations \cite{combettes1997hilbertian}. But adaptive schemes inherently imply an overhead of computation complexity which may not be compatible with certain conditions of real-time signal processing.

\subsection{Numerical experiments}

\begin{figure}
\centerline{\hbox{\scalebox{0.65}{\includegraphics{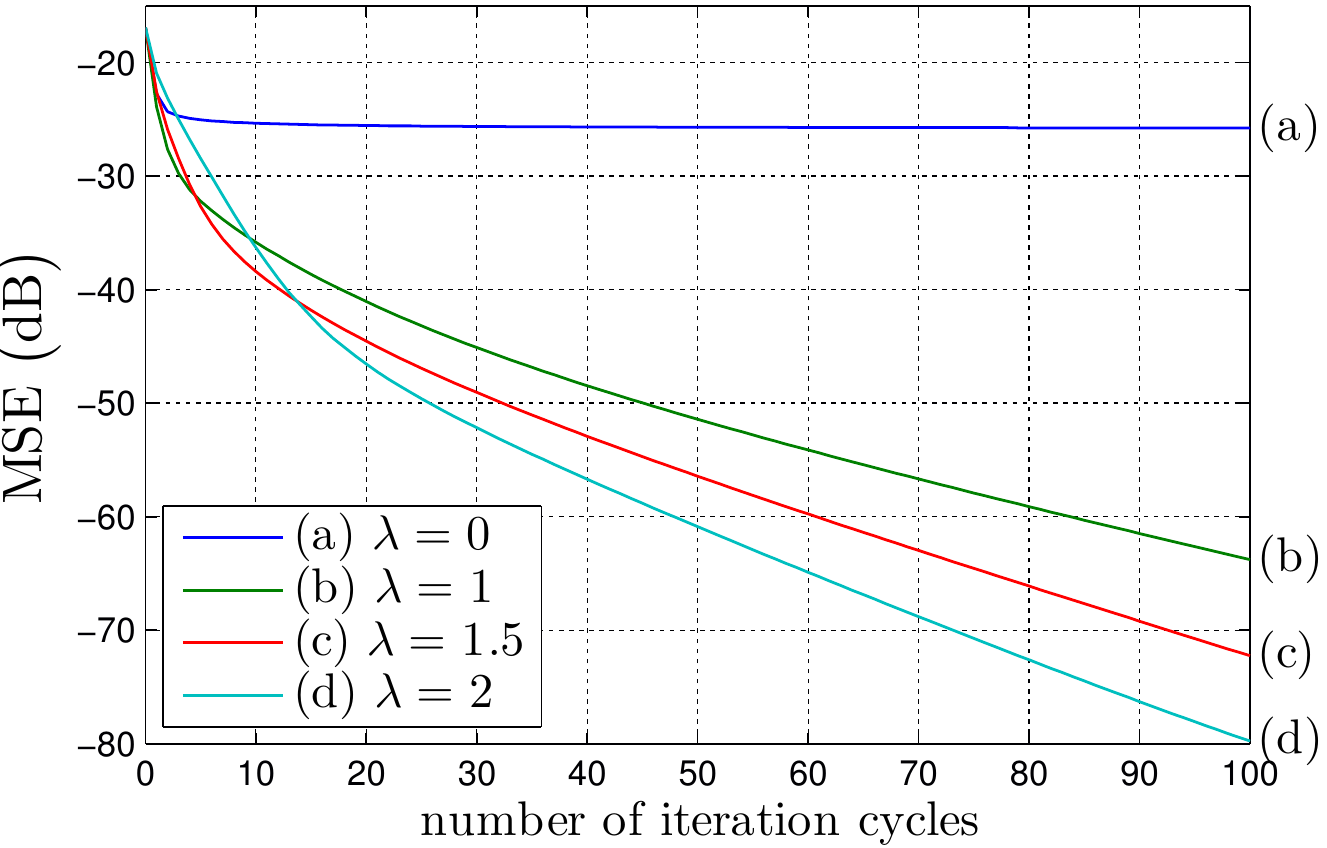}}}}
\caption{Mean squared error of iterates of the POCS method for the reconstruction of input signals from their extrema.}\label{fig:simu2_stat2}
\end{figure}

We show in Fig. \ref{fig:simu2_stat2} experimental results of the POCS iteration in the case of extrema sampling.
We assume that the encoder extracts from  a bandlimited signal $x(t)$ the time location $\tau_i$ and the amplitude value $\a_i$ of its $i$th local extremum.  Formally, this amounts to providing samples $\s_k$ of the type \eqref{deriv-samp} where for every $i$, $(n_{2i},t_{2i},\s_{2i})=(0,\tau_i,\a_i)$ and $(n_{2i+1},t_{2i+1},\s_{2i+1})=(1,\tau_i,0)$, since the derivative of $x(t)$ at its extrema locations is 0. We recall that $\s_k$ is of the form \eqref{sample} where $h_k(t)$ is given by \eqref{hk-deriv}.
In the experiment, we work with randomly generated bandlimited inputs that are periodic over an interval of length 41, assuming the Nyquist period is 1, and have exactly $N_\e=36$ extrema over one period to obtain precise statistics. To achieve such a high number of extrema, we set the input spectrum to be linearly increasing with the frequency within the bandwidth.  For each input $x(t)$, we perform the POCS iteration of \eqref{POCS-rel} with  $k\up{n}:=n~\mod~2N_\e$ (cyclic control), $\lambda\up{n}:=1$ for $n$ even, and $\lambda\up{n}:=\lambda$ for $n$ odd and some constant $\lambda$. We recall that $P^\lambda_k$ is given by \eqref{PSk} and \eqref{rel-P}. Taking $u\up{0}(t)$ to be the bandlimited version of the linear interpolation of the extrema, we measure the relative error $\|u\up{n}\!-x\|^2/\|x\|^2$ at the beginning of each iteration cycle, i.e., when $n$ is a multiple of $2N_\e$, average this over 500 inputs, and report the result in Fig. \ref{fig:simu2_stat2} versus the number of iteration cycles for various choices of the constant $\lambda$.

The case $\lambda=0$ shown in curve (a) amounts to using the extrema only as point samples and ignoring their 0-derivative property. This is a situation of sub-Nyquist sampling since only 34 samples are available for 41 Nyquist periods.  As expected, the error cannot tend to 0.  With $\lambda=1$, the POCS iteration amounts to the unrelaxed version of  \eqref{POCS}, with $2{\times}34=68$ generalized samples. This is way above the Nyquist rate and yields the MSE decay of curve (b). As seen with curves (c) and (d), using a relaxation coefficient $\lambda\in(1,2]$ accelerates the convergence.

\section{Orthogonal sampling kernel functions}\label{sec:ortho-POCS}

In the example of integrate-and-fire sampling in Section \ref{subsub:IFS}, we saw that the samples $(\s_k)\inZ$ yield the generic form of \eqref{sample} with functions $(h_k(t))\inZ$ of non-overlapping time supports as seen in \eqref{hk-int}. Thus, $(h_k(t))\inZ$ is orthogonal in $L^2(\RR)$. Back in the general space setting of this paper, we show in this section that having an orthogonal family of sampling kernel functions $(h_k)_{k\in\Z}$ in $\scH$ allows a simpler version of POCS iteration for signal reconstruction, with a number of outstanding properties concerning convergence and implementation. All our derivations in this section allow an infinite set $\Z$.

\subsection{Special POCS configuration}\label{subsec:special}

One can always rewrite the set $\scS=\cap\inZ\scS_k$ of \eqref{int-Sk} as
\begin{equation}\label{alt-inter}
\scS=\scA\cap\scC_\vs
\end{equation}
where
\begin{equation}\label{sol-set}
\scC_\vs:=\big\{v\in\scH:\forall k\in \Z,~ \langle v,h_k\rangle=\s_k\big\}
\end{equation}
and $\vs$ symbolizes the sequence $(\s_k)\inZ$. Therefore, an alternative POCS method would be the iteration of
\begin{equation}\label{x-iter1}
u\up{n+1}=P_\scA P_{\scC_\vs}u\up{n},\qquad n\geq0.
\end{equation}
But like $P_\scS$, one has in general difficult access to the projection $P_{\scC_\vs}$. Now, the outstanding contribution of having an orthogonal family of sampling kernels $(h_k)\inZ$ is that $P_{\scC_\vs}$ becomes directly accessible with the following closed form expression.
\line
\begin{proposition}
Assuming that $(h_k)\inZ$ is orthogonal in $\scH$,
\begin{equation}\label{PCs}
\forall u\in\scH,\qquad P_{\scC_\vs}u=u+\smallsum{k\in\Z}\frac{\s_k-\langle u,h_k\rangle}{\| h_k\|^2}\, h_k.
\end{equation}
\end{proposition}

\begin{IEEEproof}
For any given $j\in\Z$, $\langle h_j,h_k\rangle$ is equal to $\|h_j\|^2$ when $k=j$ and 0 otherwise. Calling $w$ be the right hand side of \eqref{PCs}, we then obtain
$\langle w,h_j\rangle=\langle u,h_j\rangle+(\s_j-\langle u,h_j\rangle)=\s_j$ for any $j\in\Z$. So $w\in\scC_\vs$. Let $v\in\scC_\vs$. For any $k\in\Z$, $\langle w{-}v,h_k\rangle=\s_k-\s_k=0$. Meanwhile, it is clear that $u-w$ is a linear combination of $(h_k)\inZ$. So $\langle u{-}w,w{-}v\rangle=0$ for any $v\in\scC_\vs$. Thus, $w=P_{\scC_\vs}u$.
\end{IEEEproof}
\ppnoi
The above proof was mainly an algebraic verification. Here are some additional explanations on the convergence and meaning of the sum in \eqref{PCs}. This result is equivalent to
\begin{equation}\label{PCs2}
\forall u\in\scH,\qquad P_{\scC_\vs} u=u+\ssum{k\in \Z}\big(\hat\s_k-\langle u,\hat h_k\rangle\big)\,\hat h_k.
\end{equation}
where
\begin{equation}\label{norm-hs}
\hat h_k:=h_k/\|h_k\|\quad\mbox{and}\quad\hat\s_k:=\s_k/\|h_k\|,\quad k\in\Z.
\end{equation}
But as $\hat\s_k=\langle x,\hat h_k\rangle$, then
$$\ssum{k\in \Z}\big(\hat\s_k-\langle u,\hat h_k\rangle\big)\,\hat h_k=\ssum{k\in \Z}\langle x{-}u,\hat h_k\rangle\,\hat h_k.$$
Since $(\hat h_k)\inZ$ is {\em orthonormal} in $\scH$, not only this sum is convergent, but it is also precisely the orthogonal projection of $x-u$ onto the closed space linearly spanned by $(\hat h_k)\inZ$.

The estimation error $\|u\up{n}\!-x\|$ is strictly decreasing as long as $u\up{n}\notin\scA\cap\scC_\vs$. In all cases,
\begin{equation}\label{x-inf2}
\lim_{n\rightarrow\infty}u\up{n}=u\up{\infty}=P_{\scA\cap\scC_\vs}\,u\up{0}.
\end{equation}
Contrary to the iteration of  \eqref{POCS}, this convergence is {\em unconditional} when $\scA\cap\scC_\vs\neq\emptyset$ even when $\Z$ is infinite.
Assuming that $u\up{0}$ is chosen in $\scA$, the iterates $u\up{n}$ of \eqref{x-iter1} remain in $\scA$ for all $n\geq0$. It then follows from \eqref{PCs2} and \eqref{x-iter1} that
\begin{equation}\label{PAPCs}
\forall u\in\scA,\quad P_\scA P_{\scC_\vs}u=u+\ssum{k\in \Z}\big(\hat\s_k-\langle u,\hat h_k\rangle\big)\,\widetilde{\hat h}_k
\end{equation}
where $\widetilde{\hat h}_k:=P_\scA\hat h_k$ according to the notation of \eqref{tilde-notation}.
\ppnoi
{\it Remark:} Using \eqref{basic-eq} and \eqref{norm-hs}, the projection $P_k u$ of \eqref{PSk} can be put for all $u\in\scA$ in the form
$$P_k u=u+\midfrac{\s_k-\langle u,h_k\rangle}{\|\widetilde h_k\|^2}\,\widetilde h_k=u+\midfrac{\hat\s_k-\langle u,\hat h_k\rangle}{\mu_k}\,\widetilde{\hat h}_k$$
where $\mu_k:=\|\widetilde h_k\|^2/\|h_k\|^2$. Thus, $P_\scA P_{\scC_\vs}$ yields the equivalent expression
\begin{equation}\label{parralel-form}
\forall u\in\scA,\qquad P_\scA P_{\scC_\vs}u=u+\ssum{k\in \Z}\mu_k(P_k u-u).
\end{equation}
This is the parallel-projection form of \eqref{PKmu} with $K=\Z$ and coefficients $\mu_k$ that are in $(0,1]$ by Bessel's inequality.
\label{}

\subsection{Linear operator presentation}

The POCS iteration limit of \eqref{x-inf2} is guaranteed when $\scA\cap\scC_\vs$ is non-empty. In practice however, the sampling sequence $\vs=(\s_k)\in\Z$ is not exactly obtained by \eqref{sample} due to inherent noise. The problem is that the iteration \eqref{x-iter1} is consistent with \eqref{sample}. As a result, $\scA\cap\scC_\vs$ might be empty and it is no longer clear what is the behavior of $u\up{n}$ at the limit. As $u\up{n}$ from \eqref{x-iter1} can be presented as
$$u\up{n}=(P_\scA P_{\scC_\vs})^n u\up{0},\qquad n\geq0,$$
the goal is to study the dependence of $\lim_{n\rightarrow\infty}(P_\scA P_{\scC_\vs})^n u\up{0}$ with both the initial estimate $u\up{0}$ and the sequence $\vs$, regardless of how $\vs$ has been produced.
This analysis is facilitated as $P_\scA P_{\scC_\vs}$ is affine, i.e., linear plus a fixed translation. Its linear part is extracted by defining the following linear operators
\begin{equation}\label{op}
\begin{array}[t]{rcl}
\hS:~\,\scA & \rightarrow &\ell^2(\Z)\\
 u & \mapsto & \big(\langle u,\hat h_k\rangle\big)\inZ
 \end{array}
\!\!\mbox{and}~~
\begin{array}[t]{rcl}
\hS^*:\,\ell^2(\Z)\!\!& \rightarrow & \scA\\
 \vc & \mapsto & \smallsum{k\in\Z}\c_k\,\widetilde{\hat h}_k
\end{array}
\end{equation}
where $\ell^2(\Z)$ is the space of square-summable sequences $\vc=(\c_k)\inZ$. By Bessel's inequality, we have not only the guarantee that the range of $\hS$ is in $\ell^2(\Z)$, but also that $\hS$ is bounded of norm $\| \hS\|\leq1$. Then \eqref{PAPCs} can be presented as
\begin{equation}\label{PAPCs-op}
\forall u\in\scA,\quad P_\scA P_{\scC_\vs}u=u+\hS^*(\hvs-\hS u)
\end{equation}
where
\begin{equation}\label{hS}
\hvs:=(\hat\s_k)\inZ=(\s_k/\|h_k\|)\inZ.
\end{equation}
We call $\hvs$ the normalized version of the sampling sequence $\vs$.
The notation $\hS^*$ has been used because it is exactly the adjoint of $\hS$. To see this, note first that $\langle u,\hat h_k\rangle=\langle u,\widetilde{\hat h}_k\rangle$ for all $u\in\scA$ from \eqref{basic-eq}. Then, denoting the canonical inner product of $\ell^2(\Z)$ by $\langle\cdot,\cdot\rangle_2$, we have for any $u\in\scA$ and $\vc\in\ell^2(\Z)$,
\begin{equation}\label{adjoint-prop0}
\langle \hS u,\bc\rangle_2=
\smallsum{k\in\Z}\langle u,\widetilde{\hat h}_k\rangle\c_k
=\Big\langle u,\smallsum{k\in\Z}\c_k \widetilde{\hat h}_k\Big\rangle=
\langle u,\hS^*\vc\rangle.
\end{equation}
Next, as \eqref{PAPCs-rel-op} can be equivalently presented as $P_\scA P_{\scC_\vs}u=(I-\hS^*\hS)u+\hS^*\hvs$ for all $u\in\scA$, where $I$ is the identity operator, the linear part of $P_\scA P_{\scC_\vs}$ is then $I-\hS^*\hS$, which is self-adjoint.

\subsection{Theorem of POCS iteration limit}

Before finding the limit of $(P_\scA P_{\scC_\vs})^n u\up{0}$, it is first interesting to characterize the emptiness condition of $\scA\cap\scC_\vs$ in terms of the operator $\hS$. As the set $\scC_\vs$ from \eqref{sol-set} can be equivalently described as
$$\scC_\vs=\big\{v\in\scH:\forall k\in \Z,~ \langle v,\hat h_k\rangle=\hat\s_k\big\},$$
one sees that
\begin{equation}\label{ASs}
\scA\cap\scC_\vs=\big\{v\in\scA:\hS v=\hvs\big\}.
\end{equation}
Thus,
$$\scA\cap\scC_\vs=\emptyset\qquad\Leftrightarrow\qquad\hvs\notin\ran(\hS)$$
where $\ran(\hS)$ designates the range of $\hS$. The next theorem gives conditions under which the convergence of $(P_\scA P_{\scC_\vs})^n u\up{0}$ is maintained and gives two characterizations of its limit.
\line
\begin{theorem}\label{theo1}
Assume that $\ran(\hS)$ is closed. Then, for any $u\up{0}\in\scA$ and any sequence $\vs$ such that $\hvs\in\ell^2(\Z)$,
\begin{align}
\lim_{n\rightarrow\infty}(P_\scA P_{\scC_\vs})^n u\up{0}
&=P_{\scA\cap\scC_\bvs}\,u\up{0}\label{eq:theo1-0}\\[-0.5ex]
&=u\up{0}+\hS^\dagger(\hvs-\hS u\up{0})\label{eq:theo1}
\end{align}
where
\begin{itemize}
\item $\bvs$ is the sequence whose normalized version $\hbvs$ is the orthogonal projection of $\hvs$ onto $\ran(\hS)$,
\item $\hS^\dagger$ is the Moore-Penrose pseudo-inverse of $\hS$ \cite{Luenberger69}.
\end{itemize}
\end{theorem}
\ppnoi
By definition,
\begin{eqnarray}
&\hS^\dagger\vc:=\argmin\limits_{v\in\scM_\vc}\|v\|,\qquad\vc\in\ell^2(\Z)\label{S+}\\
\mbox{where}&&\quad\quad\nonumber\\
&\scM_\vc:=\big\{v\in\scA:\|\hS v-\vc\|_2\mbox{ is minimized}\big\}\label{Mc}
\end{eqnarray}
and $\|\cdot\|_2$ is the canonical norm of $\ell^2(\Z)$. We prove this theorem in Appendix \ref{app:theo1}. When $\hvs\in\ran(\hS)$, then $\bvs=\vs$, so \eqref{eq:theo1-0} is just the same as \eqref{x-inf2} and the proof focuses on \eqref{eq:theo1}. When $\hvs\notin\ran(\hS)$, it then shows that \eqref{eq:theo1-0} and \eqref{eq:theo1} hold based on the fundamental property of adjoint operators
\begin{equation}\label{adjoint-prop}
\null(\hS^*)=\ran(\hS)^\perp
\end{equation}
where $\null(\hS^*)$ is the null space of $\hS^*$. This in fact directly results from \eqref{adjoint-prop0}.
The characterization \eqref{eq:theo1} was previously proved in \cite{Thao21a,Thao22a}, but only for $u\up{0}=0$ and with operators of different normalization settings.

The closed property of $\ran(\hS)$ is necessary for $\bvs$ to be defined. In the pseudo-inversion interpretation, this property is necessary for $\|\hS  v-\vc\|_2$ to have a minimum with respect to $v\in\scA$. It is by default realized when $\Z$ is finite, which is always the case in practice. In this case also, $\ell^2(\Z)$ is just $\RR^\Z$ equipped with its Euclidean norm, so any sequence $\vs=(\s_k)\inZ$ is allowed by the theorem.

\subsection{Dual interpretation of POCS iteration limit and impact}

Theorem \ref{theo1} gives two interpretations of the POCS iteration limit $u\up{\infty}$.

\subsubsection{Set theoretic interpretation}

When $\scA\cap\scC_\vs$ is non-empty, we know from \eqref{x-inf2} that $u\up{\infty}$ is the element of this set that is closest to $u\up{0}$. When $\scA\cap\scC_\vs$ is empty, the POCS iteration behaves as if this set has been replaced by $\scA\cap\scC_\bvs$ where $\bvs$ is the closest possible sequence to $\vs$ after normalization while ensuring a non-empty set $\scA\cap\scC_\bvs$. Due to the normalization, the distance between $\bvs$ and $\vs$ that is minimized is the square root of
$\sum\inZ|\bar\s_k-\s_k|^2/\|h_k\|^2$.

\subsubsection{Operator theoretic interpretation}

From an operator-theoretic viewpoint, \eqref{sample} tells us that $\hvs=\hS x$. When $u\up{0}$ is chosen to be 0, it follows from \eqref{eq:theo1} that $u\up{\infty}=\hS^\dagger\hvs$. Thus, the POCS method coincides with the standard procedure of solving a linear equation by pseudo-inversion. A consequence of interest is the dependence of $u\up{\infty}$ with sampling noise. In practice, the samples are more generally of the form  $\s_k=\langle x,h_k\rangle+\e_k$ where $\ve=(\e_k)\inZ$ is some unknown error sequence. In this case,
$$\hvs=\hS x+\hve\qquad\mbox{and}\qquad u\up{\infty}=\hS^\dagger\hvs
=\hS^\dagger\hS x+\hS^\dagger\hve$$
by linearity of $S^\dagger$, where $\hve$ is the normalized version of $\ve$. While $\hS^\dagger\hS x$ is the error-free reconstruction (equal to $x$ when $\hS$ is injective), $\hS^\dagger\hve$ is the error contribution of the sampling noise to the reconstruction. Now, as shown in Appendix \ref{app:theo1}, $\hS^\dagger\hve=\hS^\dagger\hbve$, where $\hbve$ is the orthogonal projection of $\hve$ onto $\ran(\hS)$. By Bessel's inequality, $\|\hbve\|_2\leq\|\hve\|_2$ (with a strict inequality when $\hve\notin\ran(\hS)$). Thus, pseudo-inversion has a sampling-noise filtering effect. Meanwhile, as $\hbve\in\ran(\hS)$, $\hbve$ cannot be distinguished from the normalized sampling sequence of an actual signal in $\scA$. So the error component $\hbve$ is irreversible.

\subsection{Discrete-time implementation of iteration}\label{subsec:discrete}

As a result of \eqref{PAPCs-op}, the iteration of \eqref{x-iter1} is explicitly
\begin{equation}\label{x-iter3}
u\up{n+1}=u\up{n}+\hS^*(\hvs-\hS u\up{n}),\qquad n\geq0
\end{equation}
assuming $u\up{0}\in\scA$. In practice, this is an iteration of continuous-time functions. Now,  whether the space $\scA$ has a countable dimension {\em or not}, there is a way to obtain $u\up{n}$ by a pure discrete-time iteration in $\ell^2(\Z)$. Let $\vs=(\s_k)\inZ$ be the given sequence of samples and $\hvs$ be its normalized version as defined by \eqref{hS}. For any given initial estimate $u\up{0}$, consider the system iteration
\begin{subequations}\label{sys}
\begin{align}
\vc\up{n+1}&=\vc\up{n}+(\hvs-\hS u\up{0})-\hS\hS^*\vc\up{n}\label{discr-iter}\\
u\up{n}&=u\up{0}+\hS^*\vc\up{n}\label{xc}
\end{align}
\end{subequations}
for $n\geq0$ with $\vc\up{0}:=0$, the zero vector of $\ell^2(\Z)$. We have
\begin{align*}
u\up{n+1}&=u\up{0}+\hS^*\vc\up{n+1}\\
&=u\up{0}+\hS^*\vc\up{n}+\hS^*\big(\hvs-\hS(u\up{0}{+}\hS^*\vc\up{n})\big)\\
&=u\up{n}+\hS^*(\hvs-\hS u\up{n}).
\end{align*}
Thus $u\up{n}$ reproduces the recursion of \eqref{x-iter3} for the given initial estimate $u\up{0}$. The outstanding contribution of \eqref{sys} is that $\vc\up{n}\in\ell^2(\Z)$ for all $n\geq0$. So \eqref{discr-iter} is a pure discrete-time iteration. In it, $\hvs-\hS u\up{0}$ is a fixed sequence in $\ell^2(\Z)$ which just needs to be computed once. Meanwhile, $\hS^*\hS$ can be seen as a square matrix of coefficients
\begin{equation}\label{inner-mat}
\hS\hS^*=\begin{bmatrix}\langle\widetilde{\hat h}_{k'},\hat h_k\rangle\end{bmatrix}_{(k,k')\in\Z\times\Z}
\end{equation}
which also needs to be determined once.
In practice, if one aims at the estimate $u\up{m}$, then one only needs to iterate \eqref{discr-iter} alone $m$ times, and then perform the conversion of \eqref{xc} from discrete time to Hilbert space function, only {\em once} at $n=m$.

\section{Multi-channel orthogonal sampling}\label{sec:multi-TEM}

In this section, we illustrate the theoretical power of Section \ref{sec:ortho-POCS} by applying it to the POCS method used in \cite{Adam20,Adam21} for the multi-channel time-encoding system of Fig. \ref{fig:Karen}\footnote{The letters $`x'$ and $`y'$ from \cite{Adam20,Adam21} have been interchanged in Fig. \ref{fig:Karen} to be compatible with the notation of the present article.}. In this process, we reformalize these references, while uncovering fundamental and practical consequences behind their method.

\subsection{System description}

The time-encoding system of \cite{Adam20,Adam21}  assumes that the source signals are multidimensional bandlimited functions
$$\by(t)=(y^1(t),{\cdots},y^M(t))\in\scB^N.$$
Next, instead of sampling the functions $y^i(t)$ individually, the system first expands $\by(t)$ into a redundant representation
$$\bx(t):=\A\by(t),\qquad t\in\RR$$
where $\A$ is a full rank $M{\times}N$ matrix with $M\geq N$. The signal $\bx(t)$ is thus of the form
$$\bx(t)=(x^1(t),{\cdots},x^M(t)),\qquad t\in\RR.$$
Each component $x^j(t)$ is then processed through an ASDM-based time-encoding machine  which outputs a sequence of spikes located at some increasing time instants $(t_j^i)_{j\in\Z_i}$, where $\Z_i$ is some index set of consecutive integers. From the derivations of \cite{Lazar04}, this provides the knowledge of successive integral values
\begin{equation}\label{sij}
\s_{i,j}:=\int_{t^i_{j-1}}^{t^i_j}x^i(t)\,\dif t,\qquad j\in\Z_i.
\end{equation}
The work of \cite{Adam20,Adam21} uses a POCS iteration to retrieve $\bx(t)$, before $\by(t)$ is recovered with the relation
$$\forall t\in\RR,\qquad\by(t)=\A^+\bx(t)$$
where $\A^+$ is the matrix pseudo-inverse of $\A$.

\subsection{Signal and system formalization}\label{subsec:sys-formal}

The POCS method of \cite{Adam20,Adam21} formally takes place in the Hilbert space $\scH:=(L^2(\RR))^M.$
Each element $\bu\in\scH$ is a function of time $\bu=\bu(t)=(u^1(t),{\cdots},u^M(t))$. The canonical inner product of $\scH$ is defined by
\begin{equation}\label{multi-inprod}
\langle\bu,\bv\rangle:={\textstyle\sum\limits_{i=1}^M}\int_\RR u^i(t)v^i(t)\dif t,\qquad\bu,\bv\in\scH.
\end{equation}
We will simply denote by $\|\cdot\|$ the norm on $\scH$ induced by $\langle\cdot,\cdot\rangle$,
The signal $\bx(t)$ to be retrieved is an element $\bx\in\scH$ that lies more specifically in the closed subspace
$$\scA:=\big\{\bv\in\scB^M:\forall t\in\RR,~\bv(t)\in\ran(\A)\big\}.$$
To be consistent with the generalized sampling presentation of \eqref{sample}, we are going to show that the samples $\s_{i,j}$ of \eqref{sij} can be formalized as
\begin{equation}\label{sij-formal}
\s_{i,j}=\big\langle \bh_{i,j},\bx\big\rangle,\qquad(i,j)\in\Z.
\end{equation}
Naturally,
$$\Z:=\big\{(i,j):i\in\M \mbox{ and } j\in\Z_i\big\}~~\mbox{where}~~
\M:=\{1,\cdots,M\}.$$
Then, \eqref{sij-formal} clearly coincides with \eqref{sij} by taking
\begin{equation}\label{hij-2}
\bh_{i,j}(t):=\big(0,{\cdots},0,h_j^i(t),0,{\cdots},0\big)
\end{equation}
where $h_j^i(t)$ is at the $i$th coordinate position and is the indicator function of the interval $[t_{j-1}^i,t_j^i]$. It is clear as a result that $(\bh_{i,j})_{(i,j)\in\Z}$ is an orthogonal family of $\scH$. In fact, to obtain this property, it is sufficient to have $$(h_j^i(t))_{j\in\Z_i} \mbox{  orthogonal in } L^2(\RR),\qquad i\in\M.$$
This is just what we are going to assume in this section, making the samples $(\s_{i,j})_{(i,j)\in\Z}$ more general than \eqref{sij}. For concise notation, we will simply write
\begin{equation}\label{hij}
\bh_{i,j}(t)=h_j^i(t)\,\ve_i
\end{equation}
where $\ve_i$ designates the $i$th coordinate vector of $\RR^M$.

\subsection{POCS iteration}

All results of Section \ref{sec:ortho-POCS} on the POCS method are applicable with the space and sampling settings of the above Section \ref{subsec:sys-formal}, with the notation change that every element of $\scH$ is symbolized by a bold face letter $\bu$ instead of $u$, and the sample indice $k$ have the form of a pair $(i,j)$ as seen in \eqref{sij-formal}. Up to these modifications, the POCS method can be implemented by iterating the discrete-time recursion of \eqref{discr-iter}, and executing \eqref{xc} only at the final iteration. These operations involve the operators $\hS$ and $\hS^*$ defined in \eqref{op}. They depend on $(\hat h_k)\inZ$ and $(\widetilde{\hat h}_k)\inZ$ which, with the present notation, are the functions
\begin{equation}\label{thhij}
\hspace{-2mm}\hat\bh_{i,j}=\bh_{i,j}/\|\bh_{i,j}\|\quad\mbox{and}\quad
\widetilde{\hat\bh}_{i,j}=P_\scA\hat\bh_{i,j},\quad(i,j)\in\Z.
\end{equation}
For their derivation, we will use for any $ u(t)\in L^2(\RR)$ the notation
\begin{equation}\label{PB}
\tilde u(t):=P_\scB u(t)=\sinc(t)*u(t)
\end{equation}
where $*$ denotes convolution and $\sinc(t):=\sin(\pi t)/(\pi t)$. We first have the following result.
\line
\begin{proposition}\label{prop:PA-mult}
Let $\bu(t):=u(t)\,\vu$ where $u(t)\in L^2(\RR)$ and $\vu\in\RR^M$.
\begin{enumerate}[label=(\roman*)]
\item If $u(t)\in\scB^\perp$ or $\vu\in\ran(\A)^\perp$, then $\bu(t)\in\scA^\perp$.\vspace{1mm}
\item $P_\scA\bu(t)=\tilde u(t)\,\P\vu$, where $\P:=\A\A^+$.
\end{enumerate}
\end{proposition}
\line
\begin{IEEEproof}
(i) Let $\bv=\bv(t)\in\scA$. By thinking of $\vu$ and $\bv(t)$ for each $t\in\RR$ as column vectors and interchanging the two summations of \eqref{multi-inprod}, we obtain
$$\langle\bu,\bv\rangle=\int_\RR u(t)\big(\vu^\top\bv(t)\big)\,\dif t.$$
Clearly, $\vu^\top\bv(t)\in\scB$. So if $u(t)\in\scB^\perp$, then $\langle\bu,\bv\rangle=0$. Meanwhile, if $\vu\in\ran(\A)^\perp$, then $\vu^\top\bv(t)=0$ for each single $t\in\RR$. Then  $\langle\bu,\bv\rangle=0$ regardless of $u(t)$. This proves (i).

(ii) Let $\bw(t):=\tilde u(t)\,\P\vu$. Its $i$th component is $w^i(t)=\tilde u(t)\,\q^i\in\scB$, where $\q^i$ is the $i$th coordinate of $\P\vu$. So $\bw(t)\in\scB^M$. Meanwhile, $\P\vu\in\ran(\A)$, so $\bw(t)\in\ran(\A)$ for each $t\in\RR$. Then, $\bw(t)\in\scA$. Next, we can write
$$\bu(t)-\bw(t)=\big(u(t){-}\tilde u(t)\big)\vu+\tilde u(t)\,(\vu{-}\P\vu).$$
While $u(t){-}\tilde u(t)\in\scB^\perp$, $\vu-\P\vu\in\ran(\A)^\perp$ because $\P$ is precisely the orthogonal projection of $\RR^M$ onto $\ran(\A)$. So $\bu(t)-\bw(t)\in\scA^\perp$ according to (i). Thus, $\bw(t)=P_\scA\bu(t)$.
\end{IEEEproof}
\ppnoi
It is clear from \eqref{hij-2} that $\|\bh_{i,j}\|$ is equal to the $L^2$-norm $\|h_j^i\|$ of $h_j^i(t)$. It then results from \eqref{hij}, \eqref{thhij} and Proposition \ref{prop:PA-mult}\,(ii) that
\begin{equation}\label{thhij-2}
\hat\bh_{i,j}(t)=\frac{ h_j^i(t)}{\|h_j^i\|}\,\ve_i\quad\mbox{and}\quad
\widetilde{\hat\bh}_{i,j}(t)=\frac{\tilde h_j^i(t)}{\|h_j^i\|}\,\P\ve_i
\end{equation}
for all $(i,j)\in\Z$.
This allows us to find the matrix $\hS\hS^*$ described in \eqref{inner-mat} with these functions. Seeing that $\hat\bh_{i,j}(t)$ and $\widetilde{\hat\bh}_{i,j}(t)$ are both of the form $u(t)\,\vu$, and finding from \eqref{multi-inprod} that
$$\big\langle u(t)\vu,v(t)\vv\big\rangle=\langle u,v\rangle\,\vu^\top\vv$$
where $\langle u,v\rangle$ is without ambiguity the inner product of $L^2(\RR)$, the  coefficients of $\hS\hS^*$ are
\vspace{-1mm}
\begin{eqnarray}\label{multi-inner}
&\displaystyle\big\langle\widetilde{\hat\bh}_{i',j'},\hat\bh_{i,j}\big\rangle=
\frac{\langle\tilde h^{i'}_{j'},h^i_j\rangle}{\|h^{i'}_{j'}\|\,\| h^i_j\|}\,\p_{ii'}\\
\mbox{where}&
\p_{ii'}:=(\P\ve_i)^\top\P\ve_{i'}=\ve_i^\top\P\!^\top\P\ve_{i'}=\ve_i^\top\P\ve_{i'}&\qquad~~\nonumber
\end{eqnarray}
by property of orthogonal projections. Then $(\p_{ii'})_{i,{i'}\in\M}$ are nothing but the entries of the matrix $\P$. We will see later on how the inner products $\langle\tilde h^{i'}_{j'},h^i_j\rangle$ can be obtained from a single variable lookup table.

\subsection{Final iterate output}

Once \eqref{discr-iter} has been iterated the desired number of times $n$, one can output the continuous-time multi-channel signal
$$\bu\up{n}(t)=\hS^*\vc\up{n}$$
from \eqref{xc}. We wish to know the explicit expression of $\hS^*\vc$ for any $\vc=(\c_{i,j})_{(i,j)\in\Z}\in\ell^2(\Z)$. It follows from \eqref{op} and \eqref{thhij-2} that
\begin{eqnarray}
&\hS^*\vc=\ssum{(i,j)\in \Z}\c_{i,j}\,\widetilde{\hat\bh}_{i,j}(t)
=\ssum{i\in \M}(\sinc(t)* c^i(t))\,\P\ve_i\nonumber\\
\lefteqn{\mbox{where}}\quad&
c^i(t):=\ssum{j\in\Z_i}\c_{i,j}\,\hat h_j^i(t).&\quad\label{ci}
\end{eqnarray}
If one needs to provide an estimate of the source signal $\by(t)$, then one naturally considers
$$\bv\up{n}(t):=\A\!^+ \bu\up{n}(t)=\A\!^+\hS^*\vc\up{n}\in\scB^N.$$
Since $\P=\A\A\!^+$ and $\A\!^+\A$ is identity, then
$\A\!^+\P\ve_i=\A\!^+\ve_i=\va^+_i$,
where $\va^+_i$ is the $i$th column vector of $\A\!^+$. We finally obtain for any $\vc\in\ell^2(\Z)$,
$$\A\!^+\hS^*\vc=\ssum{i\in \M}(\sinc(t)* c^i(t))\,\va^+_i$$
where $c^i(t)$ is given in \eqref{ci}.

Returning to the explicit case of \eqref{sij}, we mentioned in Section \ref{subsec:sys-formal} that
\begin{equation}\label{h-ASDM}
h_j^i(t)=1_{[t_{j-1}^i,t_j^i]}(t)
\end{equation}
where $1_I(t)$ is for any given interval $I$ its indicator function.
Then, $\hat h_j^i(t)=h_j^i(t)/\|h_j^i\|=1_{I_j^i}(t)/(t_j^i-t_{j-1}^i)^{1\!/2}$. So
$$c^i(t)=\ssum{j\in\Z_i}\c_j^i\;1_{I_j^i}(t)\qquad\mbox{where}\qquad
\c_j^i:=\midfrac{\c_{i,j}}{(t_j^i-t_{j-1}^i)^{1\!/2}}.$$
This is nothing but the piecewise constant function equal to $\c_j^i$ in $I_j^i$ for each $j\in\Z_i$. This is produced by analog circuits using a zero-order hold.

\subsection{Numerical experiments}\label{subsec:Karen-exp}

\begin{figure}
\centerline{\hbox{\scalebox{0.65}{\includegraphics{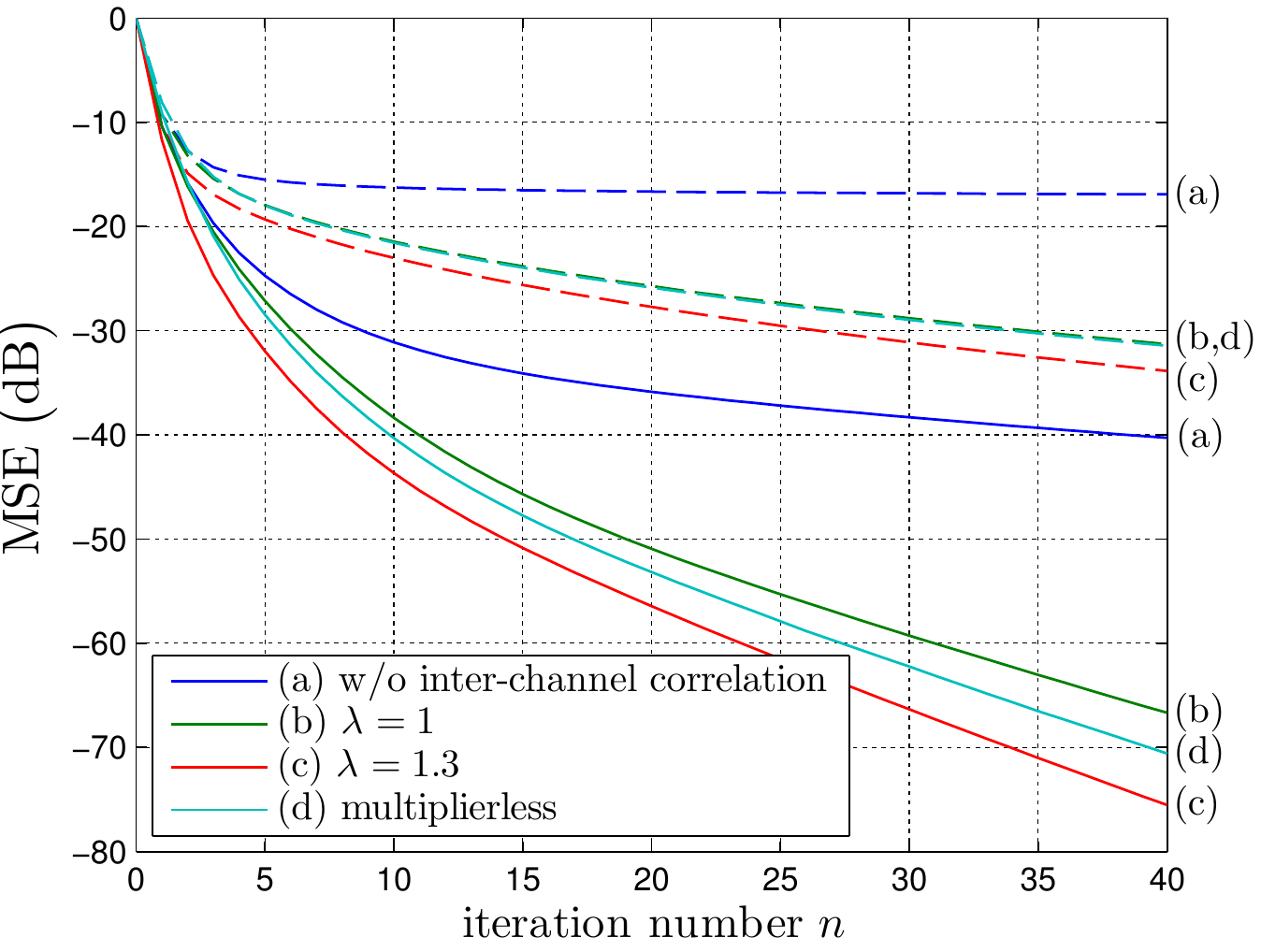}}}}
\caption{MSE results of POCS iteration for signal reconstruction from the multi-channel time encoding system of Fig. \ref{fig:Karen} with $N=2$ and $M=3$, for global oversampling ratios of $1.56$ (solid curves) and $1.49$ (dashed curves).}\label{fig:figiter5}
\end{figure}

In Fig. \ref{fig:figiter5}, we test the POCS iteration in a case where $N=2$ and $M=3$. For best uniformity between the channels, we choose a $3{\times}2$ matrix $\A$ with row vectors that form a tight frame of $\RR^2$ \cite{vetterli14}. Similarly to the experiment of Fig. \ref{fig:simu2_stat2},  the components $(y^1(t),y^2(t))$ of the source signal $\by(t)$ are bandlimited and periodic of period 61. The MSE values reported in the curves (b) of the figure are obtained by computing for a given multi-channel signal $\bx(t)=\A\by(t)$ the $n$th iterate $\bu\up{n}(t)$ of \eqref{x-iter3}, or equivalently \eqref{sys}, measuring the relative  error $\|\bu\up{n}\!-\bx\|^2/\|\bx\|^2$, and averaging this value over 100 randomly generated inputs. The solid and the dashed curves are obtained with two different densities of output samples, which can be modified by adjusting the integrator gain of the channel ASDM's. The dashed curves are obtained with an average oversampling ratio of 0.99 within each channel, while the solid curves correspond to a ratio of 1.04. Given the multi-channel configuration, this corresponds to an overall system oversampling of 1.49 and 1.56, respectively. One can see the extreme sensitivity of the MSE decay rate with the oversampling ratio. The curves (a) are obtained by omitting the redundancy between the channels in the POCS iteration, which amounts to replacing $P_\scA$ in \eqref{x-iter4} by $P_{\scB^M}$, although still applying $P_\scA$ once right before calculating the MSE. One can observe from the dashed curve (a) that the MSE stagnates to a constant. This is expected given the sub-Nyquist sampling ratio of $0.99$ in each channel. The curves (c) and (d) will be presented at the end of Sections \ref{subsec:ortho-rel} and \ref{subsec:multless}, respectively.

\section{Low complexity implementations}\label{sec:bin}

There exists also a special relaxed version for the POCS iteration of \eqref{x-iter1} for potential convergence accelerations. After describing this extended version, we show that an extra contribution of relaxation is a multiplication-free implementation of the iteration. For illustration, we give a complete description of this implementation for the multi-channel time encoding of \cite{Adam20,Adam21}. The computation of sinc-based analytical functions required for bandlimitation is implemented by simple table lookup. As is the case in practice, we assume in this section that $\Z$ is finite. In this situation, $\ell^2(\Z)$ is simply $\RR^\Z$ equipped with its Euclidean norm.

\subsection{POCS with relaxation coefficients}\label{subsec:ortho-rel}

We saw in \eqref{POCS-rel} a more general method to converge to the intersection $\scS$ using relaxation coefficients. A way to relax the iteration of \eqref{x-iter1} while keeping iterates in $\scA$ is to consider
\begin{equation}\nonumber\label{x-iter-relax}
u\up{n+1}=P_\scA P^{\lambda\up{n}}_{\scC_\vs}u\up{n},\qquad n\geq0
\end{equation}
where $(\lambda\up{n})_{n\geq0}$ is a sequence in $[\eps,2{-}\eps]$ for some $\eps\in(0,1]$ and $P^\lambda$ is defined in \eqref{rel-P}. From the expression \eqref{PCs} of $P_{\scS_\vs}$, one easily finds that
$$P^\lambda_{\scC_\vs} u=u+\lambda\,\ssum{k\in \Z}\big(\hat\s_k-\langle u,\hat h_k\rangle\big)\,\hat h_k.$$
Now, for any sequence of coefficients $\blambda=(\lambda_k)\inZ\in\RR^\Z$, consider the more general relaxed version of $P_{\scC_\vs}$ defined by
$$P^\blambda_{\scC_\vs} u:=u+\ssum{k\in \Z}\lambda_k\,\big(\hat\s_k-\langle u,\hat h_k\rangle\big)\,\hat h_k,\qquad u\in\scH.$$
It was shown in \cite{Thao21a} that the iterates of
\begin{equation}\label{x-iter4}
u\up{n+1}=P_\scA P^{\blambda\up{n}}_{\scC_\vs}u\up{n},\qquad n\geq0
\end{equation}
still converge to an element of $\scA\cap\scC_\vs$ when there exists a constant $\eps\in(0,1]$ such that $\blambda\up{n}=(\lambda\up{n}_k)\inZ$ with $\lambda\up{n}_k\in[\eps,2{-}\eps]$ for all $k\in\Z$ and $n\geq0$. Again, it is clear that
\begin{equation}\label{PAPCs-rel}
\forall u\in\scA,\qquad P_\scA P^\blambda_{\scC_\vs} u=u+\ssum{k\in \Z}\lambda_k\,\big(\hat\s_k-\langle u,\hat h_k\rangle\big)\,\widetilde{\hat h}_k.
\end{equation}
{\it Remark:} Similarly to the final remark of Section \ref{subsec:special}, one finds the equivalent expression
$$\forall u\in\scA,\qquad P_\scA P^\blambda_{\scC_\vs}u=u+\ssum{k\in \Z}\mu_k(P_k u-u).$$
with $\mu_k:=\lambda_k\|\widetilde h_k\|^2/\|h_k\|^2$. This is of the parallel-projection form of \eqref{parralel-form} but with more relaxation in the coefficients $\mu_k$.

We have tested the iterates of \eqref{x-iter-relax} in the experimental conditions of Section \ref{subsec:Karen-exp} with coefficients $\lambda_k\up{n}$ that are equal to a constant $\lambda$. The curves (c) of Fig. \ref{fig:figiter5} give the MSE results with $\lambda=1.3$, which we have found empirically to be the optimal constant for convergence acceleration.

\subsection{Unnormalized operator presentation}

As an extension to \eqref{PAPCs-op}, one obtains from \eqref{PAPCs-rel} that
\begin{equation}\label{PAPCs-rel-op}
\forall u\in\scA,\qquad P_\scA P^\blambda_{\scC_\vs} u=u+\hS^*\Lambda(\hvs-\hS u)
\end{equation}
where $\Lambda$ is the diagonal matrix of coefficients $(\lambda_k)\inZ$. Like in Section \ref{subsec:discrete}, one can again transform \eqref{x-iter4} into a discrete-time iteration of the type of \eqref{discr-iter}. But taking advantage of the presence of $\Lambda$, we propose here a different time-discretization procedure that will simplify the matrix-vector multiplication $\hS\hS^*\vc\up{n}$ of \eqref{discr-iter}. Consider the unnormalized operators
\begin{equation}\label{op2}
\begin{array}[t]{rcl}
S:~\,\scA & \rightarrow &\RR^\Z\\
 u & \mapsto & \big(\langle u,h_k\rangle\big)\inZ
 \end{array}
\!\!\mbox{and}~~
\begin{array}[t]{rcl}
S^*:\,\RR^\Z & \rightarrow & \scA\\
 \vc & \mapsto & \smallsum{k\in\Z}\c_k\,\widetilde{h}_k
\end{array}\!\!.
\end{equation}
Denoting by $\H$ the diagonal matrix of coefficients $(\|h_k\|)\inZ$, we have
$$\hS^*= S^*\H^{-1}\qquad\mbox{and}\qquad \hS=\H^{-1} S.$$
As $\hvs=\H^{-1}\vs$, \eqref{PAPCs-rel-op} becomes
\begin{align}
\forall u\in\scA,\qquad P_\scA P^\blambda_{\scC_\vs} u&=u+S^*\H^{-1}\Lambda\big(\H^{-1}\vs-\H^{-1} S u\big)\nonumber\\
&= u+ S^*\Lambda\H^{-2}(\vs- S u)\label{PAPCs-rel-op2}
\end{align}
since $\H^{-1}$ and $\Lambda\!\up{n}$ commute as diagonal matrices. This could also be directly derived from \eqref{PCs} after applying $P_\scA$ and scaling each term of the sum by $\lambda_k$.

\subsection{Alternative discrete-time implementation of iteration}

With \eqref{PAPCs-rel-op2}, the POCS iteration of \eqref{x-iter4} is explicitly
$$u\up{n+1}= u\up{n}+ S^*\Lambda\!\up{n}\H^{-2}(\vs- S u\up{n})$$
where $\Lambda\!\up{n}$ is the diagonal matrix of coefficients $(\lambda\up{n}_k)\inZ.$
Similarly to \eqref{sys}, this can be equivalently obtained by iterating
\begin{subequations}\label{sys2}
\begin{align}
\vc\up{n+1}&=\vc\up{n}+\Lambda\!\up{n}\H^{-2}\big((\vs{-}Su\up{0})-SS^*\vc\up{n}\big)\label{discr-iter2}\\
u\up{n}&=u\up{0}+S^*\vc\up{n}\label{xc2}
\end{align}
\end{subequations}
starting with $\vc\up{0}:=0$.
The discrete-time part \eqref{discr-iter2} may look complicated. But it can be equivalently implemented by the system
\vspace{-1mm}
\begin{subequations}\label{sys3}
\begin{align}
\vb\up{n}&=\Lambda\!\up{n}\H^{-2}\vr\up{n}\label{sys3a}\\
\vr\up{n+1}&=\vr\up{n}- S S^*\vb\up{n}\label{sys3b}\\
\vc\up{n+1}&=\vc\up{n}+\vb\up{n}\label{sys3c}
\end{align}
\end{subequations}
starting with
$$\vr\up{0}:=\vs-Su\up{0}\qquad\mbox{and}\qquad\vc\up{0}:=0.$$
To see this, one first needs to verify by induction from \eqref{sys3b} and \eqref{sys3c} that $\vr\up{n}=(\vs{-}Su\up{0})- S S^*\vc\up{n}$ for all $n\geq0$. Then \eqref{discr-iter2} follows from \eqref{sys3c} and \eqref{sys3a}. Again, {\em only} the discrete-time system \eqref{sys3} needs to be iterated in actual computation. The targeted estimate $u\up{n}$ is then extracted from \eqref{xc2} only {\em once}.

The system \eqref{sys3} requires the separate precomputation of the matrices
\begin{equation}\label{H2SS*}
\H^2=\mathrm{diag}(\|h_k\|^2)\inZ~~\mbox{and}~~
S S^*=\begin{bmatrix}\langle\widetilde{h}_{k'}, h_k\rangle\end{bmatrix}_{(k,k')\in\Z\times\Z}.
\end{equation}
An advantage of this reformulation is the simpler computation of $SS^*$ which no longer includes divisions of normalization. It may be argued that the normalization action is now moved to \eqref{sys3a} via the operator $\H^{-2}$. However, the next subsection actually shows an outstanding advantage of this situation.

\subsection{Multiplierless iteration}\label{subsec:multless}

Given their degree of freedom, the diagonal matrices $\Lambda\!\up{n}$ can be designed to reduce all components of $\vb\up{n}$ to mere signed powers of 2 with a computation that does not require any full multiplication or division. From \eqref{sys3a}, these components are
\begin{equation}\label{bkn-relax}
\b\up{n}_k=\lambda\up{n}_k\,\frac{\r\up{n}_k}{\|h_k\|^2},\qquad k\in\Z.
\end{equation}
Now, instead of explicitly adjusting the values of $\lambda\up{n}_k$, we propose to literally replace \eqref{sys3a} by the following component assignment
\begin{equation}\label{bkn-bin}
\b\up{n}_k:=\frac{\rho(\r\up{n}_k)}{\rho(\|h_k\|^2)},\qquad k\in\Z
\end{equation}
where
$$\rho(\r):=\sign(\r)\,\max_{2^m\leq|\r|}2^m,\qquad \r\neq0$$
with $\rho(0):=0$. When $\r\up{n}_k\neq0$, this theoretically amounts to choosing in \eqref{bkn-relax}
\begin{equation}\label{bin-relax}
\lambda\up{n}_k:=\frac{\rho(\r\up{n}_k)}{\r\up{n}_k}\,\frac{\|h_k\|^2}{\rho(\|h_k\|^2)}.
\end{equation}
This scalar can be thought of as a virtual relaxation coefficient. Meanwhile, $\b\up{n}_k$ is simply a signed power of 2. As $\rho(\r)/\r\in(\half,1]$ for any $\r\neq0$, it follows from \eqref{bin-relax} that $\lambda\up{n}_k\in(\half,2)$ (when $\r\up{n}_k=0$, one can simply think of $\lambda\up{n}_k$ as $1$). This coefficient is not rigorously in an interval of the type $[\eps,2{-}\eps]$. However, not only does convergence appear to be maintained in practice with this technique, but it is even observed to be faster than in absence of relaxation as described at the end of this subsection.
\begin{table}
\normalsize
\begin{enumerate}
\item Precompute
\begin{description}
\item $SS^*=\begin{bmatrix}\langle\widetilde{h}_{k'}, h_k\rangle\end{bmatrix}_{(k,k')\in\Z^2}$ and $\big(\rho(\|h_k\|^2)\big)\inZ$
\end{description}
\item Initialize $\vr=\vs-Su\up{0}$ and $\vc=0$
\item Repeat $n$ times
\begin{description}
\item $\vb~=~\big(\rho(\r_k)/\rho(\|h_k\|^2)\big)\inZ$
\item $\vr~\leftarrow~\vr-SS^*\vb$
\item $\vc~\leftarrow~\vc+\vb$
\end{description}
\item Output $u\up{n}=u\up{0}+S^*\vc$.
\end{enumerate}
\caption{Computation of $u\up{n}$ with multiplierless iteration\label{tab1}}
\end{table}

The determination of $\rho(\r)$ in computer arithmetic is straightforward since
it only amounts to locating the most significant digit of the binary expansion of $|\r|$ in fixed point, and is directly given by the exponent in floating point. Then, a fraction of the type $\rho(\r)/\rho(\a)$ is just a power of 2 whose exponent is the location distance between two digits in fixed point, and is obtained by a mere difference of exponents in floating point. Now, since the components of $\vb\up{n}$ are only powers of 2, the matrix-vector multiplication $S S^*\vb\up{n}$ in \eqref{sys3b} does not imply any full multiplication but only binary shifts.

We summarize in Table \ref{tab1} the complete computation of the $n$th iterate $u\up{n}$ for given sampling sequence $\vs$ and initial estimate $u\up{0}$ with the resulting multiplierless discrete-time iteration.
We have plotted in the curves (d) of Fig. \ref{fig:figiter5} the resulting  MSE of such iterates $u\up{n}$ in the experimental conditions of Section \ref{subsec:Karen-exp}. At the oversampling ratio of $1.49$, the dashed curve (d) shows no MSE degradation compared to the basic POCS iteration (dashed curve (b)). At the higher oversampling ratio of $1.56$ however, the solid curve (d) outperforms the basic POCS iteration (solid curve (b)), even though it is obtained from a computation of lower complexity.

\subsection{Table-lookup determination of $SS^*$}

In the algorithm of Table \ref{tab1}, a remaining issue is how to compute the inner products $\langle\widetilde{h}_{k'}, h_k\rangle$ involved in $SS^*$. In typical applications, these inner products do not have algebraic expressions and can only be obtained numerically. For example, when the space $\scA$ is based on bandlimitation, the sinc function is involved in the definition of $\widetilde h_{k'}$, which makes integration difficult analytically. The solution we propose is to resort to precalculated lookup tables. The major difficulty here is that $\langle\widetilde{h}_{k'}, h_k\rangle$ depends at least on two parameters, which implies the use of multidimensional tables. In the sampling case of \eqref{ASDM}, this inner product even depends on 4 parameters! However, it is shown in this case that the lookup table can be reduced to a {\em single} parameter, up to to the extra computation of 7 additions per inner product \cite{Thao21a}.

In the multi-channel system of section \ref{sec:multi-TEM}, the entries of $SS^*$ are given by \eqref{multi-inner} without the normalization coefficients , i.e.,
$$\big\langle\widetilde\bh_{i',j'},\bh_{i,j}\big\rangle=
\big\langle\tilde h^{i'}_{j'},h^i_j\big\rangle\,\p_{ii'}$$
where  $(\p_{ii'})_{i,{i'}\in\M}$ are the entries of the matrix $\P=\A\A^+$.
We show next that $\langle\tilde h^{i'}_{j'},h^i_j\rangle$ can be expressed in terms of a single-argument numerical function.
\ppnoi
\begin{proposition}\label{prop:inprod}
Assuming that $h^i_j(t)$ is given by \eqref{h-ASDM},
\begin{equation}\label{eq:inprod}
\big\langle\tilde h^{i'}_{j'},h^i_j\big\rangle=
f( T^{i,i'}_{j,{j'-1}})-f( T^{i,i'}_{j-1,{j'-1}})-f( T^{i,i'}_{j,j'})+f( T^{i,i'}_{j-1,j'})
\end{equation}
where $T^{i,i'}_{j,j'}:=t^i_j-t^{i'}_{j'}$ and
$f(t):=\int_0^t(t{-}\tau)\, \sinc(\tau)\,\dif\tau$.
\end{proposition}
\ppnoi
This was previously derived in \cite{Thao21a} in the case of a single channel. We show in Appendix \ref{app:inprod} how this derivation is extended to multiple channels. Although $\langle\tilde h^{i'}_{j'},h^i_j\rangle$ depends on the 4 time parameters $(t^i_j,t^i_{j-1},t^{i'}_{j'},t^{i'}_{j'-1})$ and is composed of 4 terms, it is the same single-argument numerical function $f(t)$ that is used. The values of this function can be stored in a lookup table. Meanwhile, the diagonal coefficients of the matrix $\H^2$ of \eqref{H2SS*} are simply
$$\|\bh_{i,j}\|^2=\|h^i_j\|^2=t^i_j-t^i_{j-1},\qquad(i,j)\in\Z$$
as a result of \eqref{h-ASDM}.

\section{Conclusion}

We have formalized the use of the POCS method for signal reconstruction from non-uniform samples by describing the most general and abstract context where this is applicable, and giving the available properties of reconstruction from this approach. On the first aspect, this method is applicable in any Hilbert space, separable or not, with samples that can be images of the input by any linear functionals. On the second aspect, the iteration of the POCS method {\em unconditional} converges to the signal that yields the same samples as the input while being closest to the initial iterate. This is true no matter how heterogenous the samples are (for example by mixing samples of different filtered versions of the input) and whether they are uniquely characteristic of the input or not (in the first case, perfect reconstruction is automatic). Before convergence, each single iteration of the POCS method has its own contribution as it guarantees an error reduction of the current estimate in the metric sense of the Hilbert space. In the second part of the paper, we have substantially pushed the analysis of the POCS method when the kernel functions of the linear functionals are orthogonal to each other. This is a case of high interest as it covers the increasingly popular time encoding by integration. In this case, one obtains additionally a parallelized version of the iteration, the exact pseudo-inversion of the linear operator that formalizes the sampling operation (including in the presence of noise), a rigorous time discretization of the iteration (applicable even in non-separable spaces) with a multiplierless implementation option that paradoxically accelerates the convergence. For demonstration, we have applied our theory to the multi-channel time-encoding system of \cite{Adam20,Adam21}, thus reformalizing this system while uncovering new consequences on it.

\appendix

\subsection{Proof of Theorem \ref{theo1}}\label{app:theo1}

Assume first that $\hvs\in\ran(\hS)$. As already mentioned in the main text, $\hbvs=\hvs$ in this case, so \eqref{eq:theo1-0} is just the same as \eqref{x-inf2}. So let us show \eqref{eq:theo1}. The set $\scM_\hvs$ of \eqref{Mc}, which is always non-empty, has in this case the simpler form
$$\scM_\hvs=\big\{ v\in\scA:\hS  v=\hvs\big\}=\scA\cap\scC_\vs$$
where the second equality was mentioned in \eqref{ASs}.
As this set is not empty, the result of \eqref{x-inf2} is applicable and gives here
$$\lim_{n\rightarrow\infty}(P_\scA P_{\scC_\vs})^n u\up{0}=u\up{\infty}=P_{\scM_\hvs}\,u\up{0}.$$
By orthogonal projection, $u\up{\infty}$ is then the element of $\scM_\hvs$ that is closest to $u\up{0}$. By translation, $u\up{\infty}-u\up{0}$ is the element of $\scM_\hvs-u\up{0}$ that is closest to $0$. Since
$$v+u\up{0}\in\scM_\hvs~~\Leftrightarrow~~\hat Sv+\hS u\up{0}=\hvs~~\Leftrightarrow~~v\in\scM_{\hvs-\hS u\up{0}},$$
then
$$u\up{\infty}-u\up{0}=\!\argmin_{v\in\scM_\hvs-u\up{0}}\|v\|=\!\argmin_{v\in\scM_{\hvs-\hS u\up{0}}}\!\|v\|=\hS^\dagger(\hvs-\hS u\up{0})$$
according to \eqref{S+}. This leads to \eqref{eq:theo1}.

Assume now the general case $\hvs\in\ell^2(\Z)$. Because $\hbvs\in\ran(\hS)$, the result of Theorem \ref{theo1} is applicable to $\hbvs$, which implies that
\begin{equation}\label{eq:theo1-hvs}
\lim_{n\rightarrow\infty}(P_\scA P_{\scC_\bvs})^n u\up{0}=P_{\scA\cap\scC_\bvs}\,u\up{0}=u\up{0}+\hS^\dagger(\hbvs-\hS u\up{0}).
\end{equation}
Let us show that
\begin{equation}\nonumber\label{RMbc}
P_\scA P_{\scC_\bvs}=P_\scA P_{\scC_\vs}\qquad\mbox{and}\qquad\hS^\dagger\hbvs=\hS^\dagger\hvs.
\end{equation}
By construction, $\hbvs-\hvs\in\ran(S)^\perp$. It then follows from \eqref{adjoint-prop} that $\hS^*(\hbvs-\hvs)=0$, or equivalently $\hS^*\hbvs=\hS^*\hvs$. It is then easy to see from \eqref{PAPCs-op} that $P_\scA P_{\scC_\bvs}=P_\scA P_{\scC_\vs}$. Then \eqref{eq:theo1-0} immediately results from the first equality of \eqref{eq:theo1-hvs}. Next, for any $u\in\scA$, $\hS u-\hbvs$ is in $\ran(\hS)$ and is therefore orthogonal to $\hbvs-\hvs$. Then, by the Pythagorean theorem
$$\|\hat Sv-\hbvs\|_2^2+\|\hbvs-\hvs\|_2^2=\|\hat Sv-\hvs\|_2^2.$$
In terms of $u\in\scA$, $\|\hat Sv-\hbvs\|_2$ is then minimized if and only if $\|\hat Sv-\hvs\|_2$ is minimized. This proves that $\scM_\hbvs=\scM_\hvs$. It immediately follows from \eqref{S+} that $\hS^\dagger\hbvs=\hS^\dagger\hvs$. We also have
$\hS^\dagger(\hbvs-\hS u\up{0})=\hS^\dagger(\hvs-\hS u\up{0})$, which can be justified by the known fact that $\hS^\dagger$ is a linear operator, or by noting directly that $\hS u\up{0}\in\ran(\hS)$. So the second equality of \eqref{eq:theo1-hvs} remains valid after replacing $\hbvs$ by $\hvs$ in it. This leads to \eqref{eq:theo1}.

\subsection{Proof of Proposition \ref{prop:inprod}}\label{app:inprod}

It follows from \eqref{h-ASDM} and \eqref{PB} that
\begin{align*}
\big\langle\tilde h^{i'}_{j'},h^i_j\big\rangle
=\int_{t^i_{j-1}}^{t^i_j}( \sinc\,{*}\,h^{i'}_{j'})(t)\dif t.
\end{align*}
Using the identity
$\int_a^b\sinc(t{-}\tau)\dif\tau=\psi(t{-}a)-\psi(t{-}b)$
where $\psi(\tau):=\int_0^\tau \sinc(s)\,\dif s$, we have
\vspace{-2mm}
$$\vspace{-2mm}
(\sinc\,{*}\,h^{i'}_{j'})(t)=\int_{t^{i'}_{j'-1}}^{t^{i'}_{j'}}\!\!\!\!\sinc(t{-}\tau)\dif\tau
=\psi(t{-}t^{i'}_{j'-1})-\psi(t{-}t^{i'}_{j'}).$$
Thus, $\displaystyle\big\langle\tilde h^{i'}_{j'},h^i_j\big\rangle=
\int_{t^i_{j-1}}^{t^i_j}\psi(t{-}t^{i'}_{j'-1})\dif t-\int_{t^i_{j-1}}^{t^i_j}\psi(t{-}t^{i'}_{j'})\dif t.$
Defining $f(t):=\int_0^t \psi(\tau)\dif\tau$, we have for any ${k'}$,
\begin{align*}
\int_{t^i_{j-1}}^{t^i_j}\psi(t{-}t^{i'}_{k'})\dif t
&=f(t^i_j{-}t^{i'}_{k'})-f(t^i_{j-1}{-}t^{i'}_{k'})\\[-1ex]
&=f( T^{i,i'}_{j,{k'}})-f( T^{i,i'}_{j-1,{k'}}).
\end{align*}
This leads to \eqref{eq:inprod}. Since $f(t)=\int_0^t\int_0^\tau \sinc(s)\,\dif s\,\dif\tau$, one obtains the expression of $f(t)$ given in Proposition \ref{prop:inprod} from the Cauchy formula for the second repeated integral of $\sinc(t)$ (derived by integration by part noting that $\sinc(\tau)=\psi'(\tau)$).

\bibliographystyle{ieeetr}

\bibliography{reference}{}

\begin{thebibliography}{10}

\bibitem{Duffin52}
R.~Duffin and A.~Schaeffer, ``A class of nonharmonic {F}ourier series,'' {\em
  Transactions of the American Mathematical Society}, vol.~72, pp.~341--366,
  Mar. 1952.

\bibitem{Yen56}
J.~L. Yen, ``On nonuniform sampling of bandwidth-limited signals,'' {\em IRE
  Trans. Circ. Theory}, vol.~CT-3, pp.~251--257, Dec. 1956.

\bibitem{Benedetto92}
J.~Benedetto, ``Irregular sampling and frames,'' in {\em Wavelets: A Tutorial
  in Theory and Applications} (C.~K. Chui, ed.), pp.~445--507, Boston, MA:
  Academic Press, 1992.

\bibitem{Feichtinger94}
H.~G. Feichtinger and K.~Gr{\"o}chenig, ``Theory and practice of irregular
  sampling,'' in {\em Wavelets: Mathematics and Applications} (J.~Benedetto,
  ed.), pp.~318--324, Boca Raton: CRC Press, 1994.

\bibitem{Marvasti01}
F.~Marvasti, {\em Nonuniform Sampling: Theory and Practice}.
\newblock New York: Kluwer, 2001.

\bibitem{Miskowicz2018}
M.~Mi\'skowicz, ed., {\em Event-Based Control and Signal Processing}.
\newblock Embedded Systems, Boca Raton, FL, USA: CRC Press, 2018.

\bibitem{Rzepka18}
D.~Rzepka, M.~Mi\'skowicz, D.~Ko\'scielnik, and N.~T. Thao, ``Reconstruction of
  signals from level-crossing samples using implicit information,'' {\em IEEE
  Access}, vol.~6, pp.~35001--35011, 2018.

\bibitem{Alexandru20}
R.~{Alexandru} and P.~L. {Dragotti}, ``Reconstructing classes of
  non-bandlimited signals from time encoded information,'' {\em IEEE
  Transactions on Signal Processing}, vol.~68, pp.~747--763, 2020.

\bibitem{Adam21}
K.~Adam, A.~Scholefield, and M.~Vetterli, ``Asynchrony increases efficiency:
  Time encoding of videos and low-rank signals,'' {\em IEEE Transactions on
  Signal Processing}, pp.~1--1, 2021.

\bibitem{Tarnopolsky22}
S.~Tarnopolsky, H.~Naaman, Y.~C. Eldar, and A.~Cohen, ``Compressed if-tem: Time
  encoding analog-to-digital compression,'' {\em arXiv preprint
  arXiv:2210.17544}, 2022.

\bibitem{Lazar04}
A.~Lazar and L.~T. T\'oth, ``Perfect recovery and sensitivity analysis of time
  encoded bandlimited signals,'' {\em IEEE Trans. Circ. and Syst.-I}, vol.~51,
  pp.~2060--2073, Oct. 2004.

\bibitem{LAZAR2005401}
A.~A. Lazar, ``Multichannel time encoding with integrate-and-fire neurons,''
  {\em Neurocomputing}, vol.~65-66, pp.~401--407, 2005.
\newblock Computational Neuroscience: Trends in Research 2005.

\bibitem{Thao22a}
N.~T. Thao, D.~Rzepka, and M.~Miśkowicz, ``Bandlimited signal reconstruction
  from leaky integrate-and-fire encoding using {POCS},'' {\em preprint
  arXiv:2201.03006}, 2022.

\bibitem{Yeh90}
S.-J. Yeh and H.~Stark, ``Iterative and one-step reconstruction from nonuniform
  samples by convex projections,'' {\em J. Opt. Soc. Am. A}, vol.~7,
  pp.~491--499, Mar 1990.

\bibitem{Thao21a}
N.~T. {Thao} and D.~{Rzepka}, ``Time encoding of bandlimited signals:
  Reconstruction by pseudo-inversion and time-varying multiplierless {FIR}
  filtering,'' {\em IEEE Transactions on Signal Processing}, vol.~69,
  pp.~341--356, 2021.

\bibitem{Combettes93}
P.~L. Combettes, ``The foundations of set theoretic estimation,'' {\em
  Proceedings of the IEEE}, vol.~81, pp.~182--208, Feb 1993.

\bibitem{Bauschke96}
H.~H. Bauschke and J.~M. Borwein, ``On projection algorithms for solving convex
  feasibility problems,'' {\em SIAM Rev.}, vol.~38, no.~3, pp.~367--426, 1996.

\bibitem{combettes1997hilbertian}
P.~Combettes, ``Hilbertian convex feasibility problem: Convergence of
  projection methods,'' {\em Applied Mathematics and Optimization}, vol.~35,
  no.~3, pp.~311--330, 1997.

\bibitem{Kaczmarz37}
S.~Kaczmarz, ``Angen\"aherte aufl\"osung von systemen linearer gleichungen,''
  {\em Bulletin de l'Acad\'emie des Sciences de Pologne}, vol.~A35,
  pp.~355--357, 1937.

\bibitem{strohmer2009randomized}
T.~Strohmer and R.~Vershynin, ``A randomized kaczmarz algorithm with
  exponential convergence,'' {\em Journal of Fourier Analysis and
  Applications}, vol.~15, no.~2, pp.~262--278, 2009.

\bibitem{Needell14}
D.~Needell and J.~A. Tropp, ``Paved with good intentions: analysis of a
  randomized block {K}aczmarz method,'' {\em Linear Algebra Appl.}, vol.~441,
  pp.~199--221, 2014.

\bibitem{Ma15}
A.~Ma, D.~Needell, and A.~Ramdas, ``Convergence properties of the randomized
  extended {G}auss-{S}eidel and {K}aczmarz methods,'' {\em SIAM J. Matrix Anal.
  Appl.}, vol.~36, no.~4, pp.~1590--1604, 2015.

\bibitem{Adam20}
K.~{Adam}, A.~{Scholefield}, and M.~{Vetterli}, ``Encoding and decoding mixed
  bandlimited signals using spiking integrate-and-fire neurons,'' in {\em
  ICASSP 2020 - 2020 IEEE International Conference on Acoustics, Speech and
  Signal Processing (ICASSP)}, pp.~9264--9268, 2020.

\bibitem{eldar2005general}
Y.~C. Eldar and T.~Werther, ``General framework for consistent sampling in
  {H}ilbert spaces,'' {\em International Journal of Wavelets, Multiresolution
  and Information Processing}, vol.~3, no.~04, pp.~497--509, 2005.

\bibitem{Adam20b}
K.~Adam, A.~Scholefield, and M.~Vetterli, ``Sampling and reconstruction of
  bandlimited signals with multi-channel time encoding,'' {\em IEEE
  Transactions on Signal Processing}, vol.~68, pp.~1105--1119, 2020.

\bibitem{chen1987analysis}
D.~Chen and J.~Allebach, ``Analysis of error in reconstruction of
  two-dimensional signals from irregularly spaced samples,'' {\em IEEE
  transactions on acoustics, speech, and signal processing}, vol.~35, no.~2,
  pp.~173--180, 1987.

\bibitem{zakhor1990reconstruction}
A.~Zakhor and A.~Oppenheim, ``Reconstruction of two-dimensional signals from
  level crossings,'' {\em Proceedings of the IEEE}, vol.~78, no.~1, pp.~31--55,
  1990.

\bibitem{Luenberger69}
D.~G. Luenberger, {\em Optimization by vector space methods}.
\newblock John Wiley \& Sons, Inc., New York-London-Sydney, 1969.

\bibitem{vetterli14}
M.~Vetterli, V.~K. Goyal, and J.~Kovacevic, {\em Foundations of signal
  processing}.
\newblock Cambridge Univ. Press, 2014.

\end{thebibliography}

\end{document}